\documentstyle[epsf,11pt]{article}
\def\to{\rightarrow}

\def\toffphi{e^+ e^- \rightarrow f \bar{f} \phi}
\def\tozphi{e^+ e^- \rightarrow Z \phi}

\def\simlt{\rlap{\lower 3.5 pt \hbox{$\mathchar \sim$}} \raise 1pt \hbox {$<$}}
\def\simgt{\rlap{\lower 3.5 pt \hbox{$\mathchar \sim$}} \raise 1pt \hbox {$>$}}
\def\epem{\ifmmode{e^+ e^-} \else{$e^+ e^-$} \fi}
\def\Im{\mathop{{\cal I}\mskip-4.5mu \lower.1ex \hbox{\it m}}}
\def\Re{\mathop{{\cal R}\mskip-4mu \lower.1ex \hbox{\it e}}}
\def\pl #1 #2 #3 {Phys.~Lett. {\bf#1}, #2 (#3)}
\def\np #1 #2 #3 {Nucl.~Phys. {\bf#1}, #2 (#3)}
\def\zp #1 #2 #3 {Z.~Phys. {\bf#1}, #2 (#3)}
\def\pr #1 #2 #3 {Phys.~Rev. {\bf#1}, #2 (#3)}
\def\prep #1 #2 #3 {Phys.~Rep. {\bf#1}, #2 (#3)}
\def\prl #1 #2 #3 {Phys.~Rev.~Lett. {\bf#1}, #2 (#3)}
\def\mpl #1 #2 #3 {Mod.~Phys.~Lett. {\bf#1}, #2 (#3)}
\def\xx #1 #2 #3 {{\bf#1}, #2 (#3)}

\font\tenrm=cmr10
\font\tenit=cmti10
\font\elevenbf=cmbx10 scaled\magstep 1
\font\elevenrm=cmr10 scaled\magstep 1
 1

\font\ninerm=cmr9

\renewenvironment{thebibliography}[1]
 { \elevenrm
   \begin{list}{\arabic{enumi}.}
    {\usecounter{enumi} \setlength{\parsep}{0pt}
     \setlength{\itemsep}{3pt} \settowidth{\labelwidth}{#1.}
     \sloppy
    }}{\end{list}}

\begin{document}
\begin{center}
\vglue 0.6cm
{  {\elevenbf        \vglue 10pt
               PROBING THE SCALAR SECTOR IN $\toffphi$}
\\}
\vglue .3cm
\baselineskip=13pt
{\tenrm M.\ L.\ STONG \\}
{\tenit
Department of Physics, University of Wisconsin, Madison, WI 53706, USA\\}
\vglue 0.1cm
{\tenrm and\\}
\vglue 0.2cm
{\tenrm K.\ HAGIWARA \\}
{\tenit Theory Group, KEK, Tsukuba, Ibaraki 305, Japan\\}
\vglue 0.5cm
{\tenrm ABSTRACT}
\end{center}
\vglue 0.2cm
{\rightskip=3pc
 \leftskip=3pc
 \tenrm\baselineskip=12pt
 \noindent
We study possible deviations from the Standard Model in the reaction
$e^+e^- \to Z\phi$, where $\phi$ denotes a spinless neutral boson.
We show how the $Z$ decay angular correlation can be used to extract
detailed information on the $\phi$ couplings, such as the parity of $\phi$,
radiatively induced form factor effects and possible CP violation in
the scalar sector.}
\vglue 0.6cm
{\elevenbf\noindent 1. Introduction and Effective Lagrangian}
\vglue 0.1cm
\baselineskip=14pt
\elevenrm
The process $e^+ e^- \to ZH$ is expected to be the best reaction to
look for the Higgs boson of mass $\simlt 2m_W$ at LEP II
and at an early stage of the next linear \epem linear
colliders$^{1}$.
Unlike in Higgs hunting at hadron colliders$^{2}$,
we expect to learn details of the Higgs boson properties and
interactions at \epem colliders.
These include the search for a deviation from the minimal one-doublet
Higgs boson model and for possible radiative effects$^{3}$.
In fact a neutral spinless boson $\phi$ which is produced via
$\epem \to Z\phi$ may not be a Higgs boson at all, but a new
type of a bound state such as a pseudo Nambu-Goldstone boson$^{4}$
of a new strong interaction with a spontaneously broken chiral
symmetry.
The particle would then be a pseudoscalar rather than a scalar.
We may even expect to observe a CP-violating interaction in the
boson sector$^{5}$.

In this paper, we study couplings of a spinless neutral boson $\phi$,
which may be a scalar, a pseudoscalar, or some mixture of the two, to
the $Z$ boson in the process $\epem \to (Z \to f \bar{f}) \phi$ by using the
angular distributions of the final-state fermions.
We consider an effective Lagrangian which contains the Standard Model
couplings of fermions to the $Z$ and $\gamma$, and study the effects
of general mass-dimension-five $\phi Z Z$ and $\phi Z\gamma$ couplings which
respect electromagnetic gauge invariance\footnotemark:
\footnotetext{\ninerm $V_{\mu \nu} = \partial_\mu V_\nu - \partial_\nu V_\mu$,
$\widetilde{V}_{\mu \nu} \equiv {1 \over 2}
\varepsilon_{\mu \nu \alpha \beta} V^{\alpha \beta}$, with the convention
$\varepsilon_{0 1 2 3} = + 1$.}
\elevenrm
\begin{equation}
{\cal L}_{ef\! f} = a_Z \, \phi Z^\mu Z_\mu + \sum_{V= Z,\gamma}
V^{\mu \nu} \bigg[ b_V \, \phi Z_{\mu \nu} +
c_V \, \left[ (\partial_\mu \phi ) Z_\nu - (\partial_\nu \phi ) Z_\mu \right]
+ \tilde{b}_V \, \phi \widetilde{Z}_{\mu \nu} \bigg].
\label{lag}
\end{equation}
The terms $a_Z$, $b_V$, and $c_V$ alone would correspond to a CP-even
scalar $\phi$, while the term $\tilde{b}_V$ alone indicates a CP-odd
pseudoscalar.  Interference of these two sets of terms
leads to CP-violating effects.

These effective interactions contribute to the process $\tozphi$.
For electron helicity $\sigma = \pm 1$ (in units of $\hbar/2$
$^{6}$)
and $Z$ polarization $\lambda$, the helicity amplitudes
for $\tozphi$ are given by:
\begin{eqnarray}
{ \cal M}_\sigma^{(\lambda = 0)} &=& \widehat{M}_\sigma^{(\lambda = 0)}
                                        \sigma \sin \theta, \nonumber \\
{ \cal M}_\sigma^{(\lambda = \pm 1)} &=& \widehat{M}_\sigma^{(\lambda = \pm 1)}
                        { 1 + \lambda \sigma \cos \theta \over \sqrt{2}},
\label{calm}
\end{eqnarray}
where
\begin{eqnarray}
\widehat{M}_\sigma^{(\lambda = 0)} &=&
        { g_Z \left( v_e + \sigma a_e \right) \over s - m_Z^2 + i m_Z \Gamma_Z}
        \bigg( 2 \sqrt{s} \, {\omega \over m_Z} ( a_Z + (s+m_Z^2) c_Z )
        + 4 s m_Z (b_Z - c_Z) \bigg)
\nonumber \\ \label{mhat0} &&
        - e \left( 2 m_Z (b_\gamma - c_\gamma)
        + 2 { \sqrt{s} \omega \over m_Z}  c_\gamma \right), \\
\widehat{M}_\sigma^{(\lambda = \pm 1)} &=&
        {g_Z \left( v_e + \sigma a_e \right) \over s - m_Z^2 +i m_Z \Gamma_Z}
        \bigg( 2 \sqrt{s} ( a_Z + (s + m_Z^2) c_Z )
        + 4 s \omega (b_Z - c_Z) -i \lambda 4 s k \tilde{b}_Z \bigg)
\nonumber \\ \label{mhat1} &&
        - e \left( 2 \omega (b_\gamma - c_\gamma)
        + 2 \sqrt{s} c_\gamma -i \lambda 2 k \tilde{b}_\gamma \right).
\label{mhat}
\end{eqnarray}
Here $\theta$ is the polar angle of the $Z$ momentum about the electron
beam direction, and $\omega$ and $k$ are the $Z$ energy and momentum
in the $e^+ e^-$ c.m.\ frame.
The couplings are denoted by $e=\sqrt{4\pi\alpha}$,
$g_Z=e/(\sin\theta_W \cos\theta_W)$,
$v_e=-1/4+\sin^2\theta_W$, and $a_e=1/4$.
The total cross section is simply obtained from the matrix elements
of Eq.~\ref{mhat}, but does not allow us to separate contributions from
the various couplings.

\vglue 0.2cm
{\elevenbf\noindent 3. Decay Angular Correlations}
\vglue 0.1cm
\baselineskip=14pt
\elevenrm
The differential cross-section for a given electron ($f$) helicity $\sigma$
($\sigma'$) is
\begin{eqnarray} \label{dsig2}
{d \sigma \left( \sigma, \sigma^\prime \right) \over d \! \cos \! \theta
                \, d \! \cos \! \hat{\theta} \, d \hat{\varphi} } =
        {1 \over 32 \pi s} \bar{\beta} \left({m_Z^2 \over s},
                                        {m_\phi^2 \over s} \right)
        {g_Z^2 (v_f+\sigma' a_f)^2 m_Z \over 64 \pi^2 \Gamma_Z}
        \left| \sum_\lambda
                {\cal M}_\sigma^\lambda (\theta)
                d_\lambda^{\sigma^\prime} (\hat{\theta},\hat{\varphi})
         \right|^2,
\end{eqnarray}
where use has been made of the limit $\Gamma_Z \ll m_Z$,
$\hat{\theta}$ and $\hat{\varphi}$ are the fermion angles in the center-of-mass
frame of the decaying $Z$, and
\begin{eqnarray}
d_{(\lambda = 0)}^{\sigma^\prime}
                \left( \hat{\theta}, \hat{\varphi} \right) &=&
            \sigma^\prime \sin \hat{\theta} \nonumber \\ \label{smalld}
d_{(\lambda = \pm 1)}^{\sigma^\prime}
                \left( \hat{\theta}, \hat{\varphi} \right) &=&
            {1 \over \sqrt{2}} (1 \pm \sigma^\prime \cos \hat{\theta} )
                        e^{\pm i \hat{\varphi}}.
\end{eqnarray}

We expand the squared matrix elements and define asymmetries:
\begin{eqnarray}
\left| \sum_\lambda {\cal M}_\sigma^\lambda (\theta)
                    d_\lambda^{\sigma^\prime} (\hat{\theta},\hat{\varphi})
\right|^2 & = &
        {\cal F}_1 (1 + \cos^2 \hat{\theta}) +
        {\cal F}_2 (1 - 3 \cos^2 \hat{\theta}) +
        {\cal F}_3 \cos \hat{\theta} +                  \nonumber \\ &&
        {\cal F}_4 \sin \hat{\theta} \cos \hat{\varphi} +
        {\cal F}_5 \sin (2 \hat{\theta}) \cos \hat{\varphi} +
        {\cal F}_6 \sin^2 \hat{\theta} \cos (2 \hat{\varphi}) + \nonumber \\ &&
        {\cal F}_7 \sin \hat{\theta} \sin \hat{\varphi} +
        {\cal F}_8 \sin (2 \hat{\theta}) \sin \hat{\varphi} +
        {\cal F}_9 \sin^2 \hat{\theta} \sin (2 \hat{\varphi}).
\label{fi} \\
A_i &=& \int_{-1}^{1} d \cos \theta { \cal F}_i /
\int_{-1}^{1} d \cos \theta { \cal F}_1,
\label{ai} \\
A_i^\prime &=& \left( \int_0^1 - \int_{-1}^{0} \right) d \cos \theta
                { \cal F}_i /
\int_{-1}^{1} d \cos \theta { \cal F}_1. \label{ai'}
\end{eqnarray}
\begin{table}[t]
\caption{\tenrm
CP and CP$\widetilde{\rm T}$$^{\,7}$ properties of squared matrix elements, and
the corresponding observable asymmetries.
CP and CP$\widetilde{\rm T}$ conservation is indicated with a $+$,
nonconservation is indicated with a $-$.
The circles indicate that observation of the asymmetry requires
identification of the charge of the final fermion $f$. The triangles indicate
that the asymmetry may be suppressed without corresponding polarization
measurements.}
\begin{center}
\begin{tabular}{l|cc|c|cc|c} \hline
\multicolumn{1}{c|}{Matrix Elements} &
\multicolumn{2}{c|}{Properties}&
\multicolumn{1}{c|}{Observables}&
\multicolumn{1}{c}{beam}&
\multicolumn{1}{c|}{$f$}&
\multicolumn{1}{c}{$f$} \\
 & CP & CP$\widetilde{\rm T}$ & & Pol. & Pol. & charge \\ \hline
$ |\widehat{M}_\sigma^+|^2 + |\widehat{M}_\sigma^-|^2
+ |\widehat{M}_\sigma^0|^2 $ & $+$ & $+$ & $\sigma_{tot}$ & - & - & - \\ \hline
$ |\widehat{M}_\sigma^0|^2 $ & $+$ & $+$ & $A_2$ & - & - & - \\ \hline
$ |\widehat{M}_\sigma^+|^2 - |\widehat{M}_\sigma^-|^2 $ &
$-$ & $-$ & $A_1^\prime$ & $ \bigtriangleup$ & - & - \\ \cline{4-7}
 & & & $A_3$ & - & $\bigtriangleup$ & $\bigcirc$ \\ \hline
$ \Re \left[ (\widehat{M}_\sigma^+ - \widehat{M}_\sigma^-)
(\widehat{M}_\sigma ^0)^\ast \right]$ & $+$ & $+$ & $A_4$ &
$\bigtriangleup$ & $\bigtriangleup$ & $\bigcirc$ \\ \cline{4-7}
 & & & $A_5^\prime$ & - & - & - \\ \hline
$ \Re\left[ (\widehat{M}_\sigma^+ - \widehat{M}_\sigma^-)
(\widehat{M}_\sigma ^0)^\ast \right]$& $-$ & $-$ &
$A_4^\prime$ & - & $\bigtriangleup$ & $\bigcirc$ \\ \cline{4-7}
 & & & $A_5$ & $\bigtriangleup$ & - & - \\ \hline
$\Re \left[ (\widehat{M}_\sigma^+)(\widehat{M}_\sigma^-)^\ast \right]$ &
$+$ & $+$ & $A_6$ & - & - & - \\ \hline
$ \Im\left[ (\widehat{M}_\sigma^+ - \widehat{M}_\sigma^-)
(\widehat{M}_\sigma ^0)^\ast \right]$& $-$ & $+$ &
$A_7$ & $\bigtriangleup$ & $\bigtriangleup$ & $\bigcirc$ \\ \cline{4-7}
 & & & $A_8^\prime$ & - & - & - \\ \hline
$ \Im \left[ (\widehat{M}_\sigma^+ + \widehat{M}_\sigma^-)
(\widehat{M}_\sigma ^0)^\ast \right]$& $+$ & $-$ &
$A_7^\prime$ & - & $\bigtriangleup$ & $\bigcirc$ \\ \cline{4-7}
 & & & $A_8$ & $\bigtriangleup$ & - & - \\ \hline
$\Im \left[ (\widehat{M}_\sigma^+)(\widehat{M}_\sigma^-)^\ast \right]$ &
$-$ & $+$ & $A_9$ & - & - & - \\ \hline
\end{tabular}
\end{center}
\end{table}
It is remarkable that for each combination of matrix elements,
one asymmetry exists which requires measurement neither of the final
fermion spin nor of its charge, and hence all the visible Z decay modes
can be used for these measurements.
We present in Table~1 the nine combinations of matrix elements which appear.
The asymmetries, averaged over electron and summed over fermion spins,
are shown in Fig.~1 as deviations from the Standard Model value
$a_Z = g_Z m_Z/2$.
In Fig.~1a, the added couplings here are $b_Z$ and $c_Z$; the Higgs is
a pure scalar in this plot.  Taking instead a pure pseudoscalar, that is,
taking only $\tilde{b}_Z$ to
be nonzero, we would find a very different result.  Then
$A_2 = A_4 = A_5^\prime = 0$, and $A_6 \simeq - 0.42$.
Fig.~1b shows a few of the CP-violating asymmetries.
Fermion angular correlations in the process
$\toffphi$ are useful in obtaining detailed information on the
CP nature of a spinless neutral particle $\phi$.
\begin{figure}
\epsfxsize=5.0in
\epsfysize=3.0in
\begin{center}
\hspace*{0in}
\epsffile{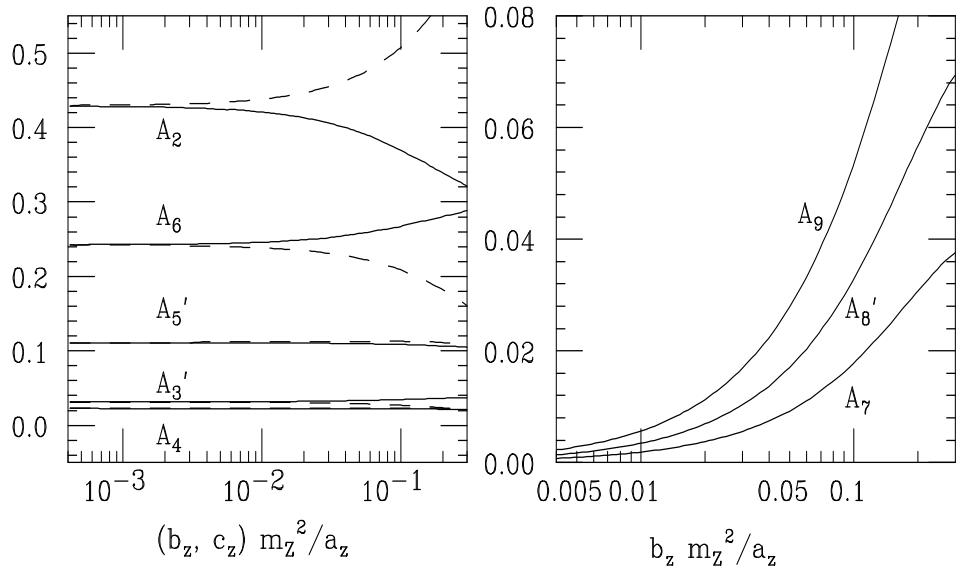}
\end{center}
\vspace{-0.25in}
\caption[asymmetries]
{\tenrm
Asymmetries of Eqs.~(8) and (9) for $m_H = 60$ GeV and $\sqrt{s} = 200$ GeV.
The solid lines indicate the effect of $b_Z$, the dashed lines give the
effect of $c_Z$.  (a) $A_2$--$A_6$ exist in the Standard Model.
(b) $A_7$--$A_9$ indicate CP violation.}
\label{fig1}
\end{figure}

\vglue 0.2cm
{\elevenbf \noindent 3. Acknowledgements \hfil}
\vglue 0.4cm
The authors wish to thank Peter Zerwas for discussions that prompted us to
investigate the problem of measuring the 'Higgs' boson couplings.
One of us (MLS) would also like to thank the NSF's Summer Institute in
Japan program. This work was funded in part by the University of
Wisconsin Research Committee with funds granted by the
Wisconsin Alumni Research Foundation, and in part by the U.S.~Department
of Energy under Contract No.~DE-AC02-76ER00881.

\vglue 0.2cm
{\elevenbf \noindent 4. References \hfil}
\vglue 0.4cm

\end{document}